\documentstyle[preprint,prb,aps]{revtex}

\begin{document}
\title{\bf $ $ \\
Accurate evaluation of the interstitial KKR-Green function}

\author{J.P. Dekker and A. Lodder} 

\address{Faculteit Natuurkunde en Sterrenkunde, Vrije Universiteit,
         De Boelelaan 1081,\\ 1081 HV Amsterdam, The Netherlands}

\author{R. Zeller}

\address{Institut f\"ur Festk\"orperforschung, Forschungszentrum J\"ulich GmbH,
         Postfach 1913,\\ 52425 J\"ulich, Germany}

\author{A.F. Tatarchenko}

\address{Institute of High Pressure Physics,
         Troitsk, Moscow region, 142092, Russia}

\date{\today}
\maketitle

\begin{abstract}
\normalsize{
It is shown that the Brillouin zone integral for the interstitial KKR-Green 
function can be evaluated accurately by taking proper care of the free-electron 
singularities in the integrand.
The proposed method combines two recently developed methods, a supermatrix
method and a subtraction method.
This combination appears to provide a major improvement compared with an
earlier proposal based on the subtraction method only.
By this the barrier preventing the study of important interstitial-like
defects, such as an electromigrating atom halfway along its jump path, can
be considered as being razed.}
\end{abstract}

\section{Introduction}
The Korringa-Kohn-Rostoker (KKR) Green function method 
\cite{korringa,kr,zelded} has proven to be
a powerful and elegant approach to calculate the electronic structure of
defects in metals \cite{bras}. 
The expressions to be evaluated are exact results of multiple 
scattering theory and the method has been applied 
successfully in calculating effects of charge transfer and lattice
distortion, both on the electronic structure \cite{bras,stephan}
and on physical quantities like the Dingle temperature \cite{molenaar}
and the effective valency of migrating atoms \cite{vanek}. 
Until recently its elegance was believed \cite{ziesche,faulkner} to arise
only after applying the muffin-tin approximation to the atomic potentials.
This amounts to an exact description of the electronic structure within
non-overlapping spheres only, being embedded in an average constant potential,
called the muffin-tin zero. Fortunately it could be proven \cite{nesb,nwtn,butl}
that the muffin-tin approximation is not necessary, opening
the possibility to do full-potential calculations in the framework
of multiple scattering theory as well.

There remains one drawback of multiple scattering theory, and this will be the 
subject of the present paper. The muffin-tin zero or free-space reference
system still appears in the expressions to be evaluated, since
free-electron poles are present in the integrand of the KKR-Green functions.
These plaguing singularities have to be handled with care. As far as
substitutional defects are concerned this problem was solved recently
\cite{zel2,deklzt} by implementing a supermatrix method.
However, for interstitial defects, such as
hydrogen in metals and an electromigrating atom halfway along its jump path,
the problem has not yet been solved. In this paper we want to present a
solution. It appears that
the supermatrix method formulated for substitutional defects can be
extended to the interstitial problem. Supplemented with a subtraction
method the expressions become manageable and evaluable to
a high degree of accuracy.

The paper is organized as follows. In section \ref{sec:relevant}
the KKR-Green function matrices of interest are defined and
the different existing computational approaches are reviewed briefly.
In section \ref{sec:supermatrix} the supermatrix method will be presented. 
In section \ref{sec:subtraction} the subtraction procedure will be described. 
In section \ref{sec:test} the subtraction method will be tested. The
paper ends with some conclusions and prospects.

\section{Relevant matrices and different approaches}
\label{sec:relevant}

In a calculation of the electronic structure of dilute alloys by use 
of the KKR-Green function technique two matrices show up \cite{dekf}, one
for defects at substitutional sites

\begin{equation} 
{\cal T}^{jj'}= 
 \frac {1} {\Omega_{BZ}} \int_{BZ} d^{3}k  e^{i{\bf k} 
\cdot{\bf R}_{jj'}}M^{-1}({\bf k}) 
\label{Tjj} 
\end{equation} 
and one for an interstitial defect

\begin{equation}
{\cal G}^{pp'}=
 \frac {1} {\Omega_{BZ}} \int_{BZ} d^{3}k  e^{i{\bf k}
\cdot{\bf R}_{pp'}}{b}^{p}({\bf k})M^{-1}({\bf k})
{b}^{p'T}(-{\bf k}).
\label{Gpp}
\end{equation}
The integrals run over the Brillouin zone ($BZ$) with volume 
${\Omega_{BZ}}$. A lattice vector ${\bf R}_{j}$ is denoted by a label $j$ and 
${\bf R}_{jj'}$ stands for the difference vector between the sites $j$ and 
$j'$. Arbitrary sites, including non-lattice sites,
are indicated by a label ${p}$. All matrices are a function of 
the energy $E$ and carry (suppressed) angular momentum labels
$L$, which stands for $(lm)$.
Both integrands contain the inverse of
the KKR matrix $M({\bf k})$ given by

\begin{equation}
M({\bf k})=t^{-1} - b({\bf k})
\label{Mk}
\end{equation}
in which the matrix $t$ expresses the scattering properties of a host atom, and
for spherical scatterers is equal to $-\sin \delta \,{e}^{i \delta}$, $\delta$
being scattering phase shifts to be labeled by the angular momentum label $l$.
The matrix $b({\bf k})$ follows from the matrix $b^{p}({\bf k})$, defined by

\begin{equation}
b^{p}({\bf k})=
\sum_{j}B^{pj}e^{-i{\bf k}\cdot{\bf R}_{pj}},
\label{bkp}
\end{equation}
after taking for ${p}$ a lattice site label. The matrix $b^{p}({\bf k})$ is
the Fourier transform of the free space propagation matrix element $B^{pj}$,
given by

\begin{equation}
B^{pj}_{LL'}=
4\pi i^{l-l'-1}\sum_{L''}i^{l''}C_{LL'L''}h^{+}_{L''}({\bf R}_{pj}),
\label{eq:B}
\end{equation}
in which $C_{LL'L''}$ are Gaunt coefficients and 
$h^{+}_{L}({\bf r}) = h^{+}_l(\kappa r)Y_{L}({\hat r})$. Real
spherical harmonics $Y_L(\hat{r})$ are used, $h^{+}_l$ are spherical Hankel
functions and $\kappa = \sqrt{E}$. The electronic structure of the metallic
host follows from the condition

\begin{equation}
\det M({\bf k})=0,
\label{kkr}
\end{equation}
which is a basic result of multiple scattering theory.

The free electron singularities are hidden in the matrix $b^{p}({\bf k})$.
They are readily made explicit by
writing down its reciprocal space representation

\begin{equation}
{b}^{p}_{LL'}({\bf k}) =
\sum_{{n}} e^{i{\bf K}_{n}\cdot{\bf R}_{pj}}
\frac {F_{L}({\bf k+K}_{n})F_{L'}^{*}({\bf k+K}_{n})} {({\bf k+K}_{n})^2-E^{+}}
+ i \hspace{.5mm}  \delta_{pj} \delta_{LL'} \frac{h^{+}_{l}
(\kappa x)}{j_{l}(\kappa x)},
\label{bkp2}
\end{equation}
in which

\begin{equation}
F_{L}({\bf k})=\sqrt {\frac{2 \Omega_{BZ}} {\pi \kappa}}i^l Y_L(\hat{k})
\frac{j_l(kx)}{j_l(\kappa x)},
\label{FL}
\end{equation}
In this expression the free electron poles at the 
energies $E=({\bf k+K}_{n})^2$ are clearly present, ${\bf K}_{n}$ denoting
a reciprocal lattice vector. The latter equality in fact defines
the so-called free electron sphere. Note, that for $p$ being a lattice
site the exponential factor reduces to unity and that only for that case the
second term contributes. The $j_l$ are spherical Bessel functions. As usual
in Green function treatments, the energy carries an infinitesimally positive
imaginary part, which is indicated by ${E^{+}}$. At this point we want to
remark, that the formalism discussed in this paper is currently \cite{bras}
applied at complex energies as well. For that slight changes in the
notation are required. However, in all calculations one has to approach the
real-energy axis somewhere, so that the pole problem shows up anyhow. It is
clear that the matrix $M({\bf k})$ also contains the free-electron
singularities. At the ${\bf k}$ points
defined by Eq. (\ref{kkr}), which pertain to the electronic structure, its
determinant value is zero, while at the free electron sphere it is singular.

Looking at the integrand of the matrix ${\cal G}^{pp'}$ it is seen, that it
is full of singularities. Free electron singularities are
present in the matrices ${b}^{p}({\bf k})$ and ${b}^{p'T}(-{\bf k})$.
These are partially cancelled by those in the matrix $M({\bf k})$, but
this matrix itself introduces poles corresponding to the electronic
structure of the metal regarding the condition Eq. (\ref{kkr}). 

In order to state the problem in actual calculations clearly we expand a little
upon it. A popular way to evaluate the matrices ${\cal T}^{jj'}$ and 
${\cal G}^{pp'}$ is using a subdivision of the Brillouin zone into
microvolumes, cubes or tetrahedrons, going back to Gilat and 
Raubenheimer \cite{gilat}, Jepsen and Andersen \cite{jepsen} 
and Lehmann and Taut \cite{lehtaut}.
The required matrix inversion is performed by using its 
eigenvalues $\lambda$ and eigenvectors $V$,

\begin{equation}
M^{-1}_{LL'} =\left(V\lambda^{-1}V^{\dagger}\right)_{LL'}=\sum_{{q}}
V_{Lq} \frac{1}{\lambda_{q}}V^{*}_{L'q} 
\label{Mvec}
\end{equation}
as suggested by Lasseter and Soven \cite{soven} and elaborated by Coleridge et al. 
\cite{col82}. The matrix ${\cal T}^{jj'}$ is the simpler one and therefore has been 
calculated most intensively. This explains the availability of quite
exhaustive studies of substitutional alloys and the relative lack of results for
interstitial alloys \cite{ellial}. 

As an introduction we concentrate on the different
evaluation methods for ${\cal T}^{jj'}$. These methods can be distinguished
by tracing the influence of the infinitesimally positive imaginary part
added to the energy ${E}$. Working out this influence explicitly a real and 
imaginary part of the integral in ${\cal T}^{jj'}$ comes out according 
to the well-known equality
 
\begin{equation}
 \frac{1}{x^{+}} =  {\cal P} \frac{1}{x} - {\pi {i} \delta (x)}
\label{xm1}
\end{equation}
The imaginary delta function part leads to a reduction of the Brillouin
zone integral to an integral over a constant energy surface. If one
is interested in properties at the Fermi energy ${E}_{F}$ that integral
runs over the Fermi surface. The tetrahedron grid reduces to a grid of
triangles over the constant energy surface. The real principal value
part remains. In practice it is evaluated along two different lines.
The most straightforward line is to evaluate the corresponding
Brillouin zone integral explicitly. This is achievable if one needs 
${\cal T}^{jj'}$ at the Fermi energy only, because quite a fine
grid of ${\bf k}$ points is required. Coleridge et al. \cite{col82} proposed to
use a weighted grid, having a finer subdivision of tetrahedra at
the singular surfaces, the Fermi surface and the free electron sphere,
and that is the way it is applied \cite{molenaar2}. For electronic structure 
calculations one needs the matrix ${\cal T}^{jj'}$ at all energies,
starting at the bottom of the band and going upwards. To that end another line of 
evaluation \cite{podl} of the real part has been developed, taking advantage of
the necessity to account for a considerable energy interval by
employing a Kramers-Kronig relation. This relation expresses the
real part of ${\cal T}^{jj'}$ in terms of its imaginary part
in the form of an integral over the energy. In this way the
Brillouin zone integration is avoided at the expense of
the necessity to evaluate the imaginary part, the constant-energy surface integral,
up to relatively high energies. At the latter point some approximation has to
be made, in choosing an upper-bound cut-off energy. It is worth while
to note, that in practice in this method the free-electron problem does
not enter in the integrals, neither in the evaluation of the real part, nor
in integrating over the constant-energy surface. The integrand
of the imaginary part simply is a product of unperturbed
metallic-host wave function coefficients \cite{podl}. The free
electron singularities only complicate the finding of
the metallic bands, using Eq. (\ref{kkr}), and the corresponding
wave function coefficients.

A relatively recent development \cite{zel2,zel3} is to evaluate the full 
Brillouin zone integral, substituting ${E^{+}} = {E + i \delta}$ and 
using a small value for $\delta$. The grid of 
${\bf k}$ points to be used is not
as fine as required for the principal value part, because ${E^{+}}$ now
is a complex number, while a value of $\delta = 0.01 E$
is already small enough to meet both the $\delta \rightarrow 0$ limit and
accuracy requirements. 

All this applies as far as the pole structure of ${M^{-1}}({\bf k})$
is concerned, corresponding to the metallic electronic
structure. Now we turn to the singularities in $M({\bf k})$
due to the free-electron poles. These singularities do not look
too serious, because ${M^{-1}}({\bf k})$ approaches zero
at these points. This is so indeed, if one
follows the so-called double linear method of Coleridge et al. \cite{col82}.
In that method the full matrix ${M^{-1}}({\bf k})$  is
represented as a ratio of two functions $n({\bf k})$ and $d({\bf k})$, and
the integral ${I(T)}$ over a tetrahedron ${T}$ is written as

\begin{equation}
I(T)=\int_{T}d^3 k \frac{n({\bf k})}{d({\bf k})}=
\sum_{i=1}^{4}K_{i}(d_{1},d_{2},d_{3},d_{4})n_{i}.
\label{IT}
\end{equation}
The third member follows analytically from the second member assuming a linear
behaviour of the functions $n({\bf k})$ and $d({\bf k})$ inside the
tetrahedron. The numbers $n_{i}$ and $d_{i}$ are the values at the
four vertices. The weights $K_{i}$ are given by Oppeneer and 
Lodder \cite{oppeneer}.
The function $d({\bf k})$ is supposed to be
zero at the singularities due to the electronic structure.
In practice one takes for $d({\bf k})$ the eigenvalue of $M({\bf k})$
which becomes zero, so the one that determines the electronic
structure. Denoting this eigenvalue by $\lambda_{0}$, the 
function $n({\bf k})$ in the numerator
is equal to $\lambda_{0} \sum_{{q}} V_{Lq} V^{*}_{L'q}/{\lambda_{q}}$.
This holds for the alkali and noble metals. 
If the Fermi surface is composed of more than one sheet more than one eigenvalue
becomes zero, of course at different ${\bf k}$ points, and
a product of the corresponding eigenvalues
is used. However, in this approach the function $n({\bf k})$
is not as linear as one might wish, even if just one
eigenvalue becomes zero. In addition to smoothly behaving
eigenvalues it contains the eigenvalue representing the free-electron singularity.
Since, due to the inverse, $n({\bf k})$ at such points becomes zero the
method still works, provided a relatively dense (weighted) grid is used.

In a later development \cite{tatar1,tatar2,zel2} it was considered to apply
Eq. (\ref{IT}) not to the full matrix ${M^{-1}}({\bf k})$, but to
each term separately in the representation (\ref{Mvec}) instead. The
functions $d_{q}({\bf k})$ in the denominator are
the roots $\lambda_{q}$, and $n_{q}({\bf k}) = V_{Lq}V^{*}_{L'q}$. 
It is clear that this cannot work without
modification, due to the terms corresponding to the highly
nonlinear free-electron roots. The modification requires an innovative
handling of the free-electron singularities. It appears that the
concept of a supermatrix has to be introduced in the description \cite{tatar1,zel2}. 
This supermatrix method, which allows for a considerable reduction of
the number of ${\bf k}$ points to be used \cite{deklzt},
will be discussed in section \ref{sec:supermatrix}.

\section{Supermatrix decomposition of the integrands}
\label{sec:supermatrix}

For the sake of clarity first the
formulation for the matrix ${\cal T}^{jj'}$  will be summarized \cite{zel2,deklzt},
after which it is given for the matrix ${\cal G}^{pp'}$.
In the supermatrix method
the terms in the matrix $b({\bf k})$, which
make the matrix $M({\bf k})$ singular, are treated separately in a special
way. Suppose that a total number of $N$ reciprocal lattice vectors
contribute to the singularity in the energy range of interest. Then,
glancing at Eqs. (\ref{Mk}) and (\ref{bkp2}), it is possible to write the
matrix $M({\bf k})$ as a sum of a smooth part $M^{0}({\bf k})$ and a part
containing the $N$ possible singularities as follows
\begin{equation}
M_{LL'}({\bf k}) = M^{0}_{LL'}({\bf k})-
\sum_{{n}}^{{N}} \frac {F_{L}({\bf k+K}_{n})F_{L'}^{*}({\bf k+K}_{n})}
{({\bf k+K}_{n})^2-E^{+}}.
\label{Mmin}
\end{equation}
After defining a square diagonal matrix $D$ with elements 
$D_{n}=({\bf k+K}_{n})^2-E^{+}$ and a rectangular matrix $F$ with
elements $F_{Ln}\equiv F_{L}({\bf k+K}_{n})$, this equation can be written in 
matrix form as

\begin{equation}   
M = M^{0}-FD^{-1}F^{\dagger}.
\label{Mmin2} 
\end{equation}
This form suggests inversion using a supermatrix $A$ defined by

\begin{equation}   
A=\left[\begin{array}{cc} 
D   &   F^{\dagger} \\  
F   &   M^{0} 
\end{array}\right] 
\label{A} 
\end{equation}   
according to the Sherman-Morrison-Woodbury formula \cite{sherman}

\begin{equation}
A^{-1} =
\left[\begin{array}{cc}
D^{-1}+D^{-1}F^{\dagger}M^{-1}FD^{-1}\hspace{5mm}  &   -D^{-1}F^{\dagger}M^{-1}   \\
     -M^{-1}FD^{-1}                    &            M^{-1}
\end{array}\right].
\label{Avec}
\end{equation}
Contrary to the original matrix $M$, the supermatrix $A$ is
regular everywhere in the Brillouin zone. The free electron poles in
$M$ appear in its supermatrix representation $A$  as
free electron zeros of the matrix $D$ in the upper left corner of $A$. 
It is even so that $\det A = \det D \det M$, which is clear from the 
following equality

\begin{eqnarray}
\det A =
\det \left[\begin{array}{cc}
D   &   F^{\dagger} \\
F   &   M^{0}
\end{array}\right]
&=&
\det \left(\left[\begin{array}{cc}
1   &   0 \\
FD^{-1}   &   1
\end{array}\right]
\left[\begin{array}{cc}
D   &   F^{\dagger} \\
0   &   M^{0}-FD^{-1}F^{\dagger}
\end{array}\right] \right)\nonumber \\
&=&\det 1 \det 1 \det D \det M ,
\label{ADM}
\end{eqnarray}
by using Eq. (\ref{Mmin2}). One simple consequence is, that the
unpleasant original KKR condition Eq. (\ref{kkr}) can be
replaced by the regular one

\begin{equation}
\det A({\bf k})=0,
\label{kkr2}
\end{equation}
This difference between $A$ and $M$ is crucial regarding the accuracy of the integration.
According to Eq. (\ref{Avec}) the wanted inverted matrix $M^{-1}$ is given simply by
the lower right block of the supermatrix $A^{-1}$. In the inversion of $A$, to
be achieved in a way similar to Eq. (\ref{Mvec}), using
its eigenvalues and eigenvectors, all eigenvalues behave smoothly.
Applying the double linear method symbolized by Eq. (\ref{IT}) to
each term separately in the sum over the inverse eigenvalues,
a mesh of about 100 ${\bf k}$ points is sufficient in most self-consistent
electronic-structure calculations. The
eigenvector products $n_{q}({\bf k})$ in the numerator even behave
so smoothly, that it appears to be sufficient to use the value
of the functions $n_{q}({\bf k})$ in the middle points of the tetrahedron only. By
that the third member of Eq. (\ref{IT}) becomes proportional to that
value, while the sum over the weights $K_{i}$ reduces to
one simple weight expression. A convincing example
of the power of the supermatrix method is given in Ref.\cite {deklzt}

Now everything is ready to focus our attention to the much more
singular integrand of the interstitial KKR-Green function matrix ${\cal G}^{pp'}$. 
Fortunately a similar decomposition of that integrand can
be designed. In addition to the form (\ref{Mmin2}) for the matrix $M$
one needs the forms

\begin{equation}   
b^{p}({\bf k}) = b^{l0}+F^{p}D^{-1}F^{\dagger}.
\label{bpl0} 
\end{equation}
and

\begin{equation}    
b^{pT}(-{\bf k}) = b^{r0}+FD^{-1}F^{p \dagger}. 
\label{bpr0}  
\end{equation} 
for the two $b$ matrices in the integrand in Eq. (\ref{Gpp}).
The matrix $F^{p}$ is defined by

\begin{equation}   
{F^{p}_{Ln}} = e^{i{\bf K}_{n} \cdot{\bf R}_{p}} {F_{Ln}}
\label{Fp} 
\end{equation}
Using the form (\ref{Avec}) for the supermatrix $A^{-1}$ it is readily seen that 
the supermatrix $P$, defined by the following product
of three supermatrices

\begin{equation}   
P=\left[\begin{array}{cc} 
D   &   F^{\dagger} \\  
0   &   b^{l0}+F^{p}D^{-1}F^{\dagger}
\end{array}\right]
\left[\begin{array}{cc} 
D   &   F^{\dagger} \\   
F   &   M^{0}  
\end{array}\right]^{-1}
\left[\begin{array}{cc} 
D   &   0 \\   
F   &   b^{r0}+FD^{-1}F^{p' \dagger}
\end{array}\right], 
\label{P}
\end{equation}   
after multiplication gets a lower right block, which is precisely
the matrix product in the integrand of Eq. (\ref{Gpp}). The left and right
supermatrices in Eq. (\ref{P}) do not behave smoothly yet, but
after some rewritings, to be given in appendix \ref{appendixA}, the
supermatrix $P$ obtains the form

\begin{equation}    
P=\left[\begin{array}{cc}  
D   &   0 \\    
0   &  - F^{p}D^{-1}F^{p' \dagger} 
\end{array}\right]+
\left[\begin{array}{cc} 
0   &   0 \\   
-F^{p}   &   b^{l0}
\end{array}\right]
\left[\begin{array}{cc}  
D   &   F^{\dagger} \\   
F   &   M^{0}    
\end{array}\right]^{-1}  
\left[\begin{array}{cc}  
0   &   -F^{p' \dagger} \\     
0   &   b^{r0}
\end{array}\right]. 
\label{Pend}
\end{equation}   
The second term contains the inverse of the supermatrix $A$, now multiplied
from the left and right by a smooth matrix. Since $A^{-1}$ can
be obtained in a smooth way, the second term is easily
evaluable. Only the free electron poles
in the lower right block of the first term still require special treatment.
It can be shown, that a subtraction procedure presented
recently \cite{deklzt} allows for a quick and accurate evaluation of
that term as well. This will be the subject of the next section.

\section{The subtraction idea}
\label{sec:subtraction}

Originally \cite{deklzt} the subtraction method was designed and applied in
handling the free electron singularities in the integrand of the matrix
${\cal G}^{pp'}$ according to
Coleridge \cite{col82}, using a weighted distribution of
tetrahedra in the Brillouin zone. For the sake of clarity that approach
will be summarized first.

The idea of the
subtraction method 
is to subtract a function $f({\bf k})$ from the integrand, which is chosen
such that the integrand gets free of the poles, while the integral
over the function $f({\bf k})$ can be evaluated analytically and is added later on.
Although it can be seen by inspecting the behaviour of
the integrand of Eq. (\ref{Gpp}) at the free electron poles that the function

\begin{equation}   
- \sum_{{n}} e^{i({\bf k+K}_{n})\cdot{\bf R}_{pp'}}
\frac {F_{L}({\bf k+K}_{n})F_{L'}^{*}({\bf k+K}_{n})} {({\bf k+K}_{n})^2-E^{+}}.
\label{calp}
\end{equation}
cancels these poles in the integrand,
an explicit proof will be given in appendix \ref{appendixB}.
This function cannot be integrated analytically. However, it is
possible to manipulate the form (\ref{calp}) such, that it retains its
pole-cancelling property on the one hand and can be integrated
on the other hand.
First the functions $F_{L}({\bf k})$ given by Eq. (\ref{FL}) can be used in a
simplified form by taking the limit $k \rightarrow \kappa$, by which the Bessel
function factor reduces to unity. This step may induce some oscillations around
the free electron singularities, but at the singularity the limit holds
exactly. Another step is the introduction of an Ewald-like convergence
factor in order to improve the convergence of the summation over
reciprocal lattice vectors. The final function $f({\bf k})$ obtains
the form
\begin{equation}
f_{LL'}({\bf k}) = - \frac{2 \Omega_{BZ}} {\pi \kappa} i^{l-l'}
\sum_{{n}} e^{i({\bf k+K}_{n})\cdot{\bf R}_{pp'}}
\frac {Y_{L}({\bf k+K}_{n})Y_{L'}({\bf k+K}_{n})} {({\bf k+K}_{n})^2-E^{+}}
e^{-{(\left|{\bf  k+K}_{n}\right|-\kappa)}^{2}/\eta},
\label{fk}
\end{equation}
in which the Ewald parameter ${\eta}$ controls the convergence.
Upon integration over the Brillouin zone, the summation over reciprocal
lattice vectors leads to an integral over all ${\bf k}$ space.
The angular part of the resulting integral can be carried out
\begin{equation}
\int d{\hat k}e^{i{\bf k}\cdot{\bf R}_{pp'}} Y_{L}({\hat k})Y_{L'}({\hat k})=
4\pi\sum_{L''} i^{l''} C_{LL'L''}j_{l''}(kR_{pp'})Y_{L''}({\hat R}_{pp'}).
\label{angint}
\end{equation}
The integral over $k$ still contains a free electron pole,
but the principal value part can be evaluated using the equality
\begin{eqnarray}
{\cal P}\int_{0}^{\infty} k^2dk
\frac{j_{l}(kR_{pp'})e^{-{(k-\kappa)}^2/\eta}}{k^2-E}=&&
{\cal P}\int_{0}^{\infty}dk
\left(\frac{k^2 j_{l}(kR_{pp'})}{k+\kappa}-\frac{\kappa}{2}j_{l}(\kappa R_{pp'})\right)
\frac{e^{-{(k-\kappa)}^2/\eta}}{k-\kappa}\nonumber \\&&+
\frac{\kappa}{4}j_{l}(\kappa R_{pp'}){\cal P}\int_{0}^{\infty}2dk
\frac{e^{-{(k-\kappa)}^2/\eta}}{k-\kappa}.
\label{subtr}
\end{eqnarray}
The first integral on the right hand side is regular, while the second
integral equals the readily available exponential integral $E_1(E/\eta)$,
being defined by \cite{abramowitz}
 \begin{equation}
 E_1(x)=\int_{x}^{\infty}\frac {e^{-t}}{t}dt.
 \label{E1}
 \end{equation}

In this way the evaluation of the matrix ${\cal G}^{pp'}$ connecting two
non-lattice sites ${\bf R}_{p}$ and ${\bf R}_{p'}$ has been made possible,
although it has to be admitted that in the used weighted mesh sometimes
over 4000 ${\bf k}$ points are required. 

Now we return to the supermatrix ${P}$, Eq. (\ref{Pend}). From Eq.
(\ref{Fp}) one readily sees that the matrix $F^{p}D^{-1}F^{p' \dagger}$
in the lower right block of the supermatrix in the right hand side
of Eq. (\ref{Pend}) has precisely the form (\ref{calp}).
Note that in Eq. (\ref{calp}) the exponential factor $e^{i{\bf k}\cdot{\bf R}_{pp'}}$
in the integrand of ${\cal G}^{pp'}$, Eq. (\ref{Gpp}), has
already been included, while the lower right block of ${P}$
represents the matrix product in Eq. (\ref{Gpp}) only.
So the only real difference between the matrix function
$-F^{p}D^{-1}F^{p' \dagger}$ and Eq. (\ref{calp}) pertains to the 
summation over reciprocal lattice vectors. In Eq. (\ref{calp}) all
of them are included, while in ${P}$ only the ${N}$ poles-generating
ones occur. From above it is clear that the function $f({\bf k})$
of Eq. (\ref{fk}) has the same properties as expression Eq. (\ref{calp}) as
far as the free electron poles are concerned. It can
be concluded that precisely that function can serve in evaluating
the remaining problematic Brillouin zone integral.
The function $-F^{p}D^{-1}F^{p' \dagger}$ can be
calculated straightforwardly, and after subtraction of $f({\bf k})$
given by Eq. (\ref{fk}) it becomes smooth. 

By this the barrier in
evaluating the interstitial KKR-Green function ${\cal G}^{pp'}$
can be considered as being razed. In addition, the achievement of
the supermatrix approach, in that a coarser mesh suffices
compared with the original Coleridge approach, remains.
The subtraction method will be tested below.

\section{Test calculations.}
\label{sec:test}

In this section the accuracy of two integrals will be tested, 
both of which suffer from the presence of free electron singularities
in the integrand. The integrations will be carried out using the
Coleridge approach \cite{col82} symbolized by Eq. (\ref{IT}).
The first integral is given by the left-hand side of the following
algebraic equality

\begin{equation}
\frac{1}{\Omega_{BZ}}\int_{BZ}b_{LL'}({\bf k})
e^{i{\bf k}\cdot{\bf R}_{jj'}}d^3 k = B^{jj'}_{LL'}.
\label{intb}
\end{equation}
The $B$ matrix in the right-hand side, for different site labels defined by 
Eq. (\ref{eq:B}), is evaluated routinely up to any desired accuracy. If $j' = j$
this matrix is defined to be zero, in accordance with the exact result 
for the integral on the left-hand side.
The $b$ matrix in the integrand, being given by the matrix $b^{p}$ of
Eq. (\ref{bkp}) for $p = j$, is seen to be singular on the free electron sphere
by inspecting the alternative representation, Eq. (\ref{bkp2}).
Comparing with the matrix function  $F^{p}D^{-1}F^{p' \dagger}$ in Eq. (\ref{Pend}),
clearly the singular part of the matrix $b$ coincides with it if $p$ and $p'$
refer to lattice sites labeled by $j$ and $j'$. So the integral of the $b$ matrix
can be considered as a close test of the subtraction procedure proposed. 
Furthermore, the choice is quite natural, because
this matrix is readily available in our computer codes.

The actual tests have been done for copper at the
Fermi energy. Using a muffin tin radius of $0.65 a/2$
and a lattice constant $a$ of $6.831 \hspace{1mm} \mbox{Bohr}$,
the phase shifts $\delta_{0}$
to $\delta_{3}$ are -0.1506388, 0.0563578, -0.1491734, and 0.0010149 respectively and
$E_{F}= 0.634 \hspace{1mm} \mbox{Ry}$. In the first test only the lattice
constant and the Fermi energy enter.

In Table \ref{table:b} results obtained without and with subtraction are compared
with the exact results according to the right-hand side of Eq. (\ref{intb}).
Only some representative matrix elements are shown, indicated by $(jlm)$-labels 
in the first six columns. The seventh to tenth columns are obtained by numerical
integration, the last column gives the exact values. The first row at the top indicates
the way in which the integration is performed, and the second and third row
specify the mesh and the
denominator function $d(\bf{k})$ in Eq. (\ref{IT}) respectively.
So, the seventh column
follows from straight integration. Application of subtraction of the
function $f(\bf{k})$, defined in Eq. (\ref{fk}), is indicated explicitly.
As for the second row at the top, straightforward integration 
requires a weighted mesh, which, in the present
example, is denser near the free electron sphere, and therefore is indicated by FES.
After subtraction the smooth integrand allows for a homogeneous grid,
indicated by 0. Nevertheless the smoothness is tested by doing the same calculation
with the weighted grid. The number of $\bf{k}$ points
is given in the last row of the table. The singular surface is the free
electron sphere and the denominator function $d(\bf{k})$ required for the
straightforward integration is a product of factors
$({\bf k+K}_{n})^2-E$, as many of them that vanish somewhere in the Brillouin zone.
Interestingly, this latter product is precisely equal to $\det D$, the determinant
value of the matrix $D$ in the upper left block of the supermatrix $A$, introduced
in Eq. (\ref{A}). This is indicated in the third row at the top.
After subtraction $d(\bf{k})$ 
can be chosen freely, and a constant, indicated by 1, is used. As an
implicit test of the linearity of $\det D$, also after subtraction
a calculation is done using that function as a denominator, given in column 10.
Of course, in that case the weighted mesh FES is required again.
Comparing the 7th and 8th columns with the last column
it is seen that application of subtraction leads to a major improvement. 
Furthermore, the better result is obtained with a much coarser mesh.
Applying the finer mesh after subtraction, as shown in the
9th column, does not lead to significant changes. In addition it can
be concluded from column 10 that the linearity of $\det D$ is
sufficient.

Another test can be derived from the matrix ${\cal G}^{pp'}$, see Eq. (\ref{Gpp}),
if the arbitrary site labels are replaced by lattice site labels.
Then, because  ${b}^{j}({\bf k}) = {b}^{j'T}(-{\bf k}) = {b}({\bf k})$,
the following algebraic identity 

\begin{equation}
{\cal G}^{jj'}=
-{t}^{-1}\delta_{jj'}-B^{jj'}+ {t}^{-1} {\cal T}^{jj'} {t}^{-1}
\label{eq:singlesite}
\end{equation}
can easily be derived, using the relation (\ref{Mk}) between $M({\bf k})$
and ${b}({\bf k})$. Both matrices $\cal G$ and $\cal T$ can be
evaluated numerically only. However, while the integrand of ${\cal G}^{jj'}$
contains free-electron singularities, in the integrand of ${\cal T}^{jj'}$
they merely appear as a nonlinearity in the eigenvalues of
the matrix $M({\bf k})$, as described in section \ref{sec:relevant}.
In the tests ${\cal G}^{jj'}$ is evaluated only according to 
the original Coleridge approach \cite{col82}, including a
subtraction procedure as well, in the way the present
authors have described recently \cite{deklzt}. The matrix 
${\cal T}^{jj'}$ will be calculated by the
supermatrix method also, for which less ${\bf k}$ points
are expected to be required. So features of both the subtraction and
supermatrix methods will be illustrated by Eq. (\ref{eq:singlesite}).

In Table \ref{table:bMb} results for ${\cal G}^{jj'}$ according to
the left-hand side of Eq. (\ref{eq:singlesite}), before and after 
applying the subtraction procedure,
are compared with results according to the right-hand side. This table
is built up the same way as Table \ref{table:b}.
First we elucidate the denominator functions $d(\bf{k})$ of Eq. (\ref{IT}),
as specified in the third row at the top. As far as the poles
corresponding to the electronic structure are concerned, the
product of the vanishing eigenvalues of $M(\bf{k})$
can be taken, as described in section \ref{sec:relevant}.
$\rm{Det} \hspace{1mm} M(\bf{k})$ might work as well, being 
equal to the product of ${all}$ eigenvalues. But the latter choice
introduces a problem because of its singular (and therefore
nonlinear) behaviour at the free electron surface.
Therefore, $\det M(\bf{k})$ has to be multiplied by $\det D$.
By that additional multiplication, regarding Eq. (\ref{ADM}), the denominator
function $d(\bf{k})$ becomes equal to $\det A$, which
appears in the modified KKR condition Eq. (\ref{kkr2}). 
The free-electron singularities in the integrand of ${\cal G}^{jj'}$
require multiplication by an additional factor of $\det D$. So
for straightforward integration, see column 7, $\det A \det D$
has to be used. In that case the (weighted) mesh must be taken to
be dense near both the Fermi surface and the free electron sphere. 
This mesh is indicated by FS/FES. 
After subtraction the free-electron singular surface is absent. This allows
for the less dense mesh FS and for  the simpler denominator 
function $\det A$, column 8. To see
the influence of the density of the mesh the same calculation
is done with the denser FS/FES mesh, column 9. In addition, also in this
complicated case, the effect of the (extra) nonlinearity of the denominator 
is investigated, by doing the same calculation with the
denominator $\det A \det D$, as shown in column 10. Columns 11 to 13
give results according to the right-hand side of Eq. (\ref{eq:singlesite}),
the last two columns showing supermatrix results for two different
meshes. Contrary to the weighted meshes of the Coleridge method,
the meshes used in the supermatrix method are always uniform.

Again the results improve largely by applying subtraction. In fact,
the straightforwardly evaluated integrals can be considered as really 
bad. An obvious source of inaccuracy is the behaviour of the
denominator function $\det A \det D$ in regions where
a tetrahedron is cut by the Fermi surface
as well as the free electron sphere. Then the  product of $\det A$ and
$\det D$ does not change sign and the denominator
does not become aware of crossing a singular surface at all. This even
may happen for a noble metal. The single sheet Fermi surface
intersects the free electron sphere in the neck region.
Comparing columns 8 and 9 one sees that the denser mesh
near the free electron surface yields not too large
but yet non-negligible modifications. The addition of the extra
factor $\det D$ is not as harmless
as in the integration of $b(\bf{k})$, as can be seen from column 10.
This is just another illustration of the source of inaccuracy indicated above.
So it can be stated that it is important to use a denser mesh at
the free electron sphere, even if the integrand is not singular there. This
certainly is a demerit of the Coleridge method. Nevertheless, it
is satisfactory to see (columns 11 to 13) that, in evaluating
the ${\cal T}$ matrix, the Coleridge
method leads to the same results as the supermatrix method. In addition
the table confirms explicitly that the subtraction method,
proposed recently \cite{deklzt} for determining the full ${\cal G}$ matrix,
is reliable. 

Finally, the table shows implicitly, that
the supermatrix method for determining the ${\cal G}$ matrix
is more efficient than the subtraction method \cite{deklzt}. 
Less than 1000 $\bf{k}$ points are sufficient. The number
of 891 in the last column is a good indication for integrating
the second term in Eq. (\ref{Pend}), which contains
the same supermatrix $A^{-1}$ as the supermatrix integrand
of the matrix ${\cal T}$. The subtraction
procedure proposed for the first term, having free electron
poles only, is tested by Table \ref{table:b}. The number of
640 $\bf{k}$ points in the 8th column suffices for that term.

\section{Conclusions and prospects}
\label{conclusions}

The supermatrix method, initially proposed 
\cite{deklzt} with the aim of a
fast and accurate evaluation of the KKR-Green function ${\cal T}^{jj'}$
appearing in calculations of the electronic
structure of substitutional alloys, has been extended to the
interstitial KKR-Green function ${\cal G}^{pp'}$. A subtraction procedure
is shown to resolve the remaining problem of integrating
a function with free electron singularities only.
In evaluating the corresponding Brillouin zone integrals rather
coarse grids of less than 1000 $\bf{k}$ points can be used. Applications
to calculations of transport properties such as the effective valence
of electromigrating atoms are in progress \cite{ek2}. Study of 
the electronic structure of the largely unexplored
interstitial defects will be the subject of future investigations.

\acknowledgments
  One of us (AFT) wants to
acknowledge support from a NATO LINKAGE Grant HTECH.CRG 931184 and a Grant
N MU3000 of the International Soros Foundation. Part of this work was sponsored
by the National Computing Facilities Foundation (NCF) for the use of supercomputer
facilities, with financial support from the Netherlands Organization for
Scientific Research (NWO).

\appendix
\section{}
\label{appendixA}

We want to give a derivation of the final form Eq. (\ref{Pend}) of the 
supermatrix $P$ from its definition Eq. (\ref{P}). Using the matrix $A$
defined by Eq. (\ref{A}) we first write the matrix $P$ a little more compact
 
\begin{equation}
P=\left[\begin{array}{cc}
D   &   F^{\dagger} \\
0   &   b^{l0}+F^{p}D^{-1}F^{\dagger}
\end{array}\right]
A^{-1}
\left[\begin{array}{cc}
D   &   0 \\
F   &   b^{r0}+FD^{-1}F^{p' \dagger}
\end{array}\right].
\label{PA}
\end{equation}
Subsequently the supermatrix left of the
supermatrix $A^{-1}$ is written in the form $X^{l}A+Y^{l}$ and
the supermatrix to the right is written similarly as $A X^{r}+Y^{r}$.
In principal, the choice of the four matrices $X^{l}$, $Y^{l}$, $X^{r}$
and $Y^{r}$ is arbitrary, but a suitable one is

\begin{equation}
X^{l}=\left[\begin{array}{cc}
 1    &  0 \\
F^{p}D^{-1} & 0
\end{array}\right],\hspace{5mm}
Y^{l}=\left[\begin{array}{cc}
 0    &  0 \\
 -F^{p} &  b^{l0}
\end{array}\right],
\label{XYl}
\end{equation}

\begin{equation}
X^{r}=\left[\begin{array}{cc}
1   &   D^{-1}F^{p' \dagger} \\
0   &   0
\end{array}\right],\hspace{5mm}
Y^{r}=\left[\begin{array}{cc}
0   &   -F^{p' \dagger} \\
0   &   b^{r0}
\end{array}\right].
\label{XYr}
\end{equation}
The elaboration of Eq. (\ref{PA}) is now straightforward and ends up with the
expression
\begin{equation}
P= \left[\begin{array}{cc}
D   &   0 \\
0   &  -F^{p}D^{-1}F^{p' \dagger}
\end{array}\right]
+ \left[\begin{array}{cc}
0   &   0 \\
-F^{p}   &   b^{l0}
\end{array}\right]
A^{-1}
\left[\begin{array}{cc}
0   &   -F^{p' \dagger} \\
0   &   b^{r0}
\end{array}\right].
\label{Pend2}
\end{equation}

Regarding the definition of the supermatrix $A$, Eq. (\ref{A}), this
is precisely the required form  Eq. (\ref{Pend}) given in the main text.

\section{}
\label{appendixB}

We want to show that the function (\ref{calp}) indeed cancels the
free electron poles in the integrand of the
interstitial KKR-Green function (\ref{Gpp}). Apart from the trivial
exponential factor $e^{i{\bf k} \cdot{\bf R}_{pp'}}$, the 
integrand, being a matrix to be denoted as $a$,
can be written in the following form

\begin{equation}
a = (b^{l0} + F^p D^{-1} F^\dagger)
    (M^{0} - F D^{-1} F^\dagger)^{-1}
    (b^{r0} + F D^{-1} F^{p'\dagger}),
\label{andr1}
\end{equation}
in which Eqs. (\ref{Mmin2}), (\ref{bpl0}) and (\ref{bpr0}) are used.
First we define a matrix $c$ 

\begin{equation}
c = (M^{0} - F D^{-1} F^\dagger)^{-1}
    (b^{r0} + F D^{-1} F^{p'\dagger}),
\label{andr2}
\end{equation}
being equal to the product of two matrices in the triple product
matrix $a$. So 

\begin{equation} 
a = b^{l0}c + F^p D^{-1} F^\dagger c,
\label{andr3}
\end{equation}
while Eq. (\ref{andr2}) can be written in the form

\begin{equation}
F D^{-1} F^\dagger c = M^{0}c - b^{r0} - F D^{-1} F^{p'\dagger}.
\label{andr4}
\end{equation}
Three observations can be made. First, from its definition it is clear
that the matrix $c$ is a regular one, because the poles in
numerator and denominator cancel. Secondly, as a consequence, only the second
term in Eq. (\ref{andr3}) contains the free electron poles. Thirdly,
the left hand side of Eq. (\ref{andr4}) already has the form of
that second term. Regarding Eq. (\ref{Fp}) it becomes equal to that second term by
multiplying Eq. (\ref{andr4}) from the left with the factor
$e^{i{\bf K}_{n} \cdot{\bf R}_{p}}$. By
that the third term on the right hand side of Eq. (\ref{andr4}) obtains
the required form  $- F^{p} D^{-1} F^{p'\dagger}$. Substituting
the rewritten form of Eq. (\ref{andr4}) in Eq. (\ref{andr3}) completes
a first proof. The merit of the second proof to be given is, 
that in addition it exhibits
a nice link \cite{tatar2} between the subtraction and supermatrix methods.

To that end an auxiliary rectangular matrix $q$ is defined by the relation
 
\begin{equation}
F^\dagger c = -F^{p'\dagger} - D q
\label{andr5}
\end{equation}
by which Eq. (\ref{andr4}) reduces to

\begin{equation}
F q = - M^{0}c + b^{r0}.
\label{andr6}
\end{equation}
Now Eqs. (\ref{andr5}) and (\ref{andr6}) can be combined in a
supermatrix form

\begin{equation}
\left[\begin{array}{ll} D & F^\dagger \\ F & M_0 \end{array} \right]
\left[\begin{array}{l} q \\ c \end{array} \right] = 
\left[\begin{array}{c} -F^{p'\dagger} \\ b^{r0} \end{array} \right]
\label{andr7} 
\end{equation}   
It is seen that the supermatrix $A$, Eq. (\ref{A}), enters the formulation.

The solution of Eq. (\ref{andr7})

\begin{equation}
\left[\begin{array}{l} q \\ c \end{array} \right] = A^{-1}
\left[\begin{array}{c}  -F^{p'\dagger} \\ b^{r0} \end{array} \right] =
\left[\begin{array}{ll} A^{-1}_{11} & A^{-1}_{12}\\ A^{-1}_{21} &A^{-1}_{22} 
      \end{array} \right]
\left[\begin{array}{c}  -F^{p'\dagger} \\ b^{r0} \end{array} \right] 
\label{andr8}
\end{equation}   
requires the subblocks $A^{-1}_{ij}$ of the supermatrix $A^{-1}$, 
by definition given by

\begin{equation}
\left[\begin{array}{ll} D & F^\dagger\\ F & M_0 \end{array} \right]
\left[\begin{array}{ll} A^{-1}_{11} & A^{-1}_{12}\\ A^{-1}_{21} &A^{-1}_{22}
             \end{array} \right] =
\left[\begin{array}{ll} I & 0 \\ 0 & I \end{array} \right].
\label{andr9}
\end{equation}
We only need the two explicitly written equations

\begin{equation}
D A^{-1}_{11} + F^\dagger A^{-1}_{21} = I 
 \rule{20pt}{0pt} \text{and} \rule{20pt}{0pt}
D A^{-1}_{12} + F^\dagger A^{-1}_{22} = 0.
\label{andr10}
\end{equation}
After substitution of the matrix $c$ according to Eq. (\ref{andr8})
in Eq. (\ref{andr3}) for the matrix of interest $a$,

\begin{equation}
a = (b^{l0} + F^p D^{-1} F^\dagger)
(- A^{-1}_{21} F^{p'\dagger} + A^{-1}_{22} b^{r0}),
\label{andr11} 
\end{equation}    
and making use of the equalities (\ref{andr10}), one finds

\begin{equation}
a = - b^{l0} A^{-1}_{21} F^{p'\dagger} + b^{l0} A^{-1}_{22} b^{r0}
+ F^p A^{-1}_{11} F^{p'\dagger} - F^p A^{-1}_{12} b^{r0}
- F^p D^{-1} F^{p'\dagger},
\label{andr12}
\end{equation}
in which form once again the term containing the free
electron poles is made explicit.

Up to now the matrix $a$ figures in the subtraction method only,
standing for the singular matrix in the integrand of the 
interstitial KKR-Green function matrix ${\cal G}^{pp'}$.
It is interesting to go one step further and to write Eq. (\ref{andr12})
in the matrix form

\begin{equation}
a = [ -F^p \rule{10pt}{0pt} b^{l0} ] \,A^{-1} 
\left[\begin{array}{c} - F^{p'\dagger} \\ b^{r0} \end{array} \right]
- F^p D^{-1} F^{p'\dagger}.
\label{andr13} 
\end{equation}
In this form one recognizes the
lower right block of the supermatrix $P$, Eq. (\ref{Pend2}), which
figures in the supermatrix method. So
the supermatrix and subtraction procedures appear
to be intimately linked.

\begin{table}[t]
\caption{$B^{jj'}_{LL'}$ calculated by integration of $b(\bf{k})$ in different
ways. The last column contains its exact value.}
\begin{tabular}{cccccccccccc}
  \multicolumn{6}{c}{$B^{jj'}_{LL'}$ labels} & 
$straight$ & $f(\bf{k})$ & $f(\bf{k})$& $f(\bf{k})$ &$exact$\\
  &    &   &  &  &  & FES & 0 & FES & FES &\\
$j$ & $j'$ & $l$ & $m$ & $l'$ & $m'$ &  $\det D$&  1 & 1 &$\det D$ &\\
\hline
1 & 1  & 0 & 0 & 0  & 0  &
 0.023 & 0.000 & 0.000& 0.001  & 0 \\
  &    & 1 & 0 & 1  & 0  &
 0.023 & 0.000 & 0.000& 0.001  & 0 \\
  &    & 2 & 1 & 2  & 1  &
 0.049 & 0.001 & 0.000& 0.003  & 0 \\
  &    & 3 & 2 & 3  & 2  &
 0.223 & 0.017 & 0.006& 0.139  & 0 \\
  &    & 1 & 0 & 3  & 0  &
 0.040 & 0.000 & 0.000&-0.002  & 0 \\
2 & 1  & 0 & 0 & 0  & 0  &
 0.207 & 0.199 & 0.199& 0.205  & 0.198\\
  &    & 1 & 0 & 1  & 0  &
 0.164 & 0.172 & 0.172& 0.178  & 0.172 \\
  &    & 2 & 1 & 2  & 1  &
 0.464 & 0.470 & 0.470 & 0.477 & 0.469 \\
  &    & 3 & 2 & 3  & 2  &
2.794 &2.649 &2.647 &2.709 &2.629 \\
  &    & 1 & 0 & 3  & 0  &
 0.457 & 0.456 & 0.455 &  0.457 & 0.455 \\
\hline
  \multicolumn{6}{c} {Number of $\bf{k}$ points} &
2247&640&2247&2247& \\
\end{tabular}
\label{table:b}
\end{table} 

\begin{table}[b] 
\caption{The matrix ${\cal G}^{jj'}_{LL'}$ calculated along different lines.}
\begin{tabular}{cccccccccccccc}
  \multicolumn{6}{c}{${\cal G}^{jj'}_{LL'}$ labels}       &  $straight$ & 
 $f(\bf{k})$ & $f(\bf{k})$& $f(\bf{k})$&
${r.h.s.}$& ${r.h.s.}$&${r.h.s.}$\\
  &    &   &   &    &         & FS/FES & FS & FS/FES & FS/FES & FS/FES 
&$super$&$super$\\
 $j$ & $j'$ & $l$ & $m$ & $l'$ & $m'$    & $\det A \det D$&
$\det A$ & $\det A$ & $\det A \det D$ & $\det A$& $matrix$&$matrix$ \\
\hline
1 & 1  & 0 & 0 & 0  & 0  
& -0.289 & -0.153 & -0.150 & -0.190 & -0.149 & -0.155 & -0.155 \\
  &    & 1 & 0 & 1  & 0 
& 0.008 & 0.166 & 0.160 & 0.106 & 0.161 & 0.180 & 0.169 \\
  &    & 2 & 1 & 2  & 1 
& 0.351 & 0.538 & 0.524 & 0.461 & 0.525 & 0.551 & 0.536 \\
  &    & 3 & 2 & 3  & 2 
&2.201 &2.311 &2.366 &2.371 &2.486 & 2.278 & 2.284 \\
  &    & 1 & 0 & 3  & 0 
& 0.061 & 0.105 & 0.089 & 0.080 & 0.087 & 0.106 & 0.106 \\
2 & 1  & 0 & 0 & 0  & 0 
& 0.014 & -0.002 & -0.006 & -0.001 & 0.000 & -0.006 & -0.005 \\
  &    & 1 & 0 & 1  & 0 
&-0.151 &-0.139 &-0.137 &-0.146 &-0.132 & -0.137 & -0.138 \\
  &    & 2 & 1 & 2  & 1 
& -0.026 & -0.047 & -0.052 & -0.054 & -0.045 & -0.045 & -0.042 \\
  &    & 3 & 2 & 3  & 2 
&-0.568 &-0.488 &-0.522 &-0.448 &-0.435 & -0.595 & -0.601 \\
  &    & 1 & 0 & 3  & 0 
& -0.027 & -0.079 & -0.080 & -0.079 & -0.077 & -0.068 & -0.061 \\
\hline
  \multicolumn{6}{c}{Number of $\bf{k}$ points}  &
4123&2643&4123&4123&4123& 146 & 891 \\
\end{tabular}
\label{table:bMb}
\end{table}

\end{document}